\documentclass[aps,prl,reprint,twocolumn]{revtex4-1}
\usepackage{latexsym}
\usepackage{amsmath}
\usepackage{amssymb,braket}
\usepackage{amsfonts}
\usepackage{subfigure}
\usepackage[english]{babel}
\usepackage[cp1250]{inputenc}
\usepackage[OT4]{fontenc}
\usepackage[pdftex]{graphicx}
\usepackage{float}
\usepackage{subdepth}

\begin{document}
\title{Chirality for crooked curves}
\author{Giovanni Dietler$^1$}
%email{}
\author{Robert Kusner$^2$}
%\email{} 
\author{W\"oden Kusner$^3$}
\email{wkusner@gmail.com} 
\author{Eric Rawdon$^4$}
%email{}
\author{Piotr Szymczak$^5$}
%email{}

\affiliation{$^1$\'Ecole Polytechnique F\'ed\'erale de Lausanne, Lab Physics of Living Matter, CH-1015 Lausanne, Switzerland}
\affiliation{$^2$Dept of Mathematics, University of Massachusetts, Amherst, MA 01003, USA}
\affiliation{$^3$Dept of Mathematics, Vanderbilt University, Nashville, TN 37240, USA}
\affiliation{$^4$Dept of Mathematics, University of St.~Thomas, Saint Paul, MN 55105, USA}
\affiliation{$^5$Inst of Theoretical Physics, Faculty of Physics, University of Warsaw, Pasteura 5, 02-093, Warsaw, Poland}

\begin{abstract}
Chiral objects rotate when placed in a collimated flow or wind. 
We exploit this hydrodynamic intuition to construct a tensorial chirality measure for rigid filaments and curves. 
This tensor is trace-free, so if a curve has a right-handed twist about some axis, there is a perpendicular axis about which the twist is left-handed. 
Our measure places minimal requirements on the smoothness of the curve, hence it can be readily used to quantify chirality for biomolecules and polymers, polygonal and rectifiable curves, and other discrete geometrical structures.
\end{abstract}

\maketitle 
%%%%%%%%%%%%%%%%%%%
%
% Section I
%
%%%%%%%%%%%%%%%%%%%
Many objects around us are reflection symmetric, so that Alice Through the Looking-Glass would note no difference, while other objects---like our hands or our shoes---come in mirror pairs.
The attempt to distinguish between such pairs led to a definition of chirality, first attributed to Kelvin~\cite{Kelvin1894,kelvin1894molecular,whyte1958chirality}:
 ``I call any geometrical figure, or group of points, \emph{chiral}, and say that it has chirality, if its image in a plane mirror, ideally realized, cannot be brought to coincide with itself.'' 

Can chirality be measured or quantified?
Perhaps a regular tetrahedron, slightly perturbed, should be much less chiral than a tetrahedron with edges of significantly different length.
Several attempts to introduce scalar measures of chirality have been made. 
Some obtain a degree of chirality for a molecule by defining a distance between it and its mirror image (enantiomers)~\cite{mislow92,rassat06}.  
Other measures are based on a model of optical activity: chiral molecules rotate the polarization plane of the irradiating light~\cite{osipov95,neal03}.
Nonetheless, the use of scalar chirality measures for objects of arbitrary shape is not without problems. 
For triangles in the plane, there exist well-behaved, continuous scalar measures making any scalene triangle most chiral~\cite{Rassat2003}. 
Similar results hold for tetrahedra in space~\cite{rassat2004there}.
It is even possible to continuously deform a chiral object onto its mirror image along a path which involves \emph{only} chiral configurations; thus there \emph{cannot} be a continuous scalar measure of chirality which is positive for the ``right-handed" chiral objects, negative for ``left-handed" objects, and zero for the achiral ones~\cite{kurt1997fuzzy, ruch1972algebraic}.

As recognized by several researchers~\cite{Ferrarini1998, Efrati2014}, the problems with a scalar measure of chirality are manifestations of the fact that chirality is tensorial: an object  considered right-handed from one direction can be left-handed when regarded from another.
This is familiar to those who study minimal surfaces like the helicoid or its close relative, the double-helix ladder, which twist one way about their axes and the other way about their rulings or rungs.

\begin{figure}[H]
\centering     
\includegraphics[width=0.45\textwidth]{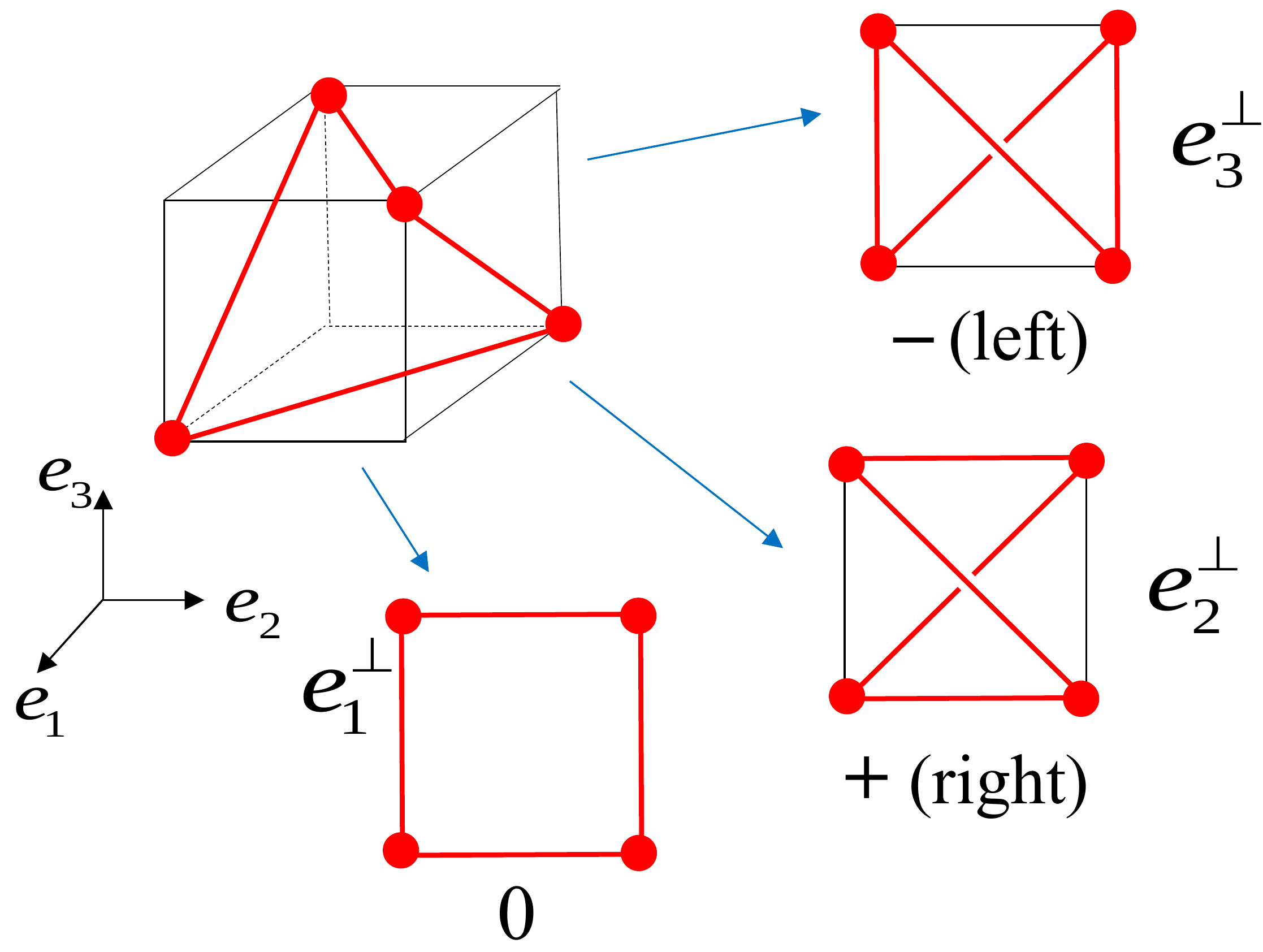}
\vspace{-0.4cm}

\caption{This space quadrilateral is achiral according to Kelvin, but it exhibits global left-and-right-handedness.}
\label{fig1}
\end{figure}

We even find evidence of tensorial chirality in a simple object which would be achiral according to Kelvin's definition: a reflection-symmetric space polygon drawn on a cube using four face-diagonals for the edges (Fig.~\ref{fig1}).
Viewing this polygon along one direction $\boldsymbol{e}_1$ we see a square, whereas viewing in the other face directions reveals a bow-tie with a right-handed or left-handed twist. 
The conclusion is that our putatively achiral object is globally ``left-and-right-handed"!

In this paper we propose a tensorial measure of chirality for wire-like objects mathematically modeled by curves---or more generally by $1$-dimensional sets of locally-finite length---in $3$-dimensional Euclidean space $\mathbb{R}^3$. In contrast to measures of chirality based on models for the optical response of molecules \cite{osipov95}, we base our chirality measure on the hydrodynamic observation that objects twist when they move relative to a fluid, and inversely, when placed in a collimated flow or \emph{wind}. 
This is seen in the spin of turbine blades, the twirl of falling tree seeds \cite{tennakone2017,varshney2011}, or the tumble of knots sedimenting in silicone oil~\cite{Weber2013}. 
For a tensor measure, the twist of an object in the wind may depend on the direction from which the wind blows, as is indeed observed~\cite{Efrati2014}.
We shall show an even stronger statement holds:  
\emph{the average of twists about any three mutually perpendicular axes}---equal to the average twist over the entire sphere of directions---\emph{must vanish.}
Consequently, an object with a right-handed twist about an axis must have a left-handed twist about some other perpendicular axis.

One can gain insight into the chirality of an object---like a wire bent into a curve of arbitrary shape \cite{Krapf2009}---sedimenting in a very viscous fluid by using the apparatus of low-Reynolds-number hydrodynamics \cite{klein1906schraubentheorie,Oseen1927,HappelBrenner1973,Kim-Karilla:1991} to approximate its equations of motion; however, the resulting formulae are rather complex and difficult to analyze. 

Since our interest is in quantifying the chirality of a curve, we propose instead a much simpler model for the interaction between the wire and wind, treating the latter as a multitude of tiny ballistic particles colliding elastically with the wire, transferring momentum. 
We adopt this model with the aim of capturing the essentials of chirality---and convincing the reader that it does so.

%%%%%%%%%%%%%%%%%%%
%
% Section II 
%
%%%%%%%%%%%%%%%%%%%
%%%%%%%%%%%%%%%%%%%
%
% Subsection A  %\subsection{Force and torque matrix densities}
%
%%%%%%%%%%%%%%%%%%%
Consider a space curve $\boldsymbol{x}=\boldsymbol{x}(s)$ in $\mathbb{R}^3$, parametrized by arclength $s$, with unit tangent vector $\boldsymbol{t}=\boldsymbol{t}(s)=\boldsymbol{x}'(s)$. 
Imagine the curve made of thin wire, placed in a wind of particles that collide with it and reflect elastically; the particles transfer momentum and exert a force on the curve.
If an incoming particle with velocity $\boldsymbol{v}$ hits the curve at $\boldsymbol{x}$, then (up to some constant scalar multiple depending on the material) the momentum transferred is the projection of $\boldsymbol{v}$ to the normal plane of the curve at $\boldsymbol{x}$.
Thus, using $\_^*$ for the usual adjoint or transpose operator, the corresponding force density along the curve is proportional to
\begin{equation}
\boldsymbol{f}=\boldsymbol{v}-\boldsymbol{t}(\boldsymbol{t}^* \boldsymbol{v} )=\mathsf{f}\boldsymbol{v} ,\end{equation}
defining the associated \emph{force density matrix} 
\begin{equation}
\mathsf{f} =  \mathsf{I}-\boldsymbol{t}\boldsymbol{t}^*\hspace{-4pt},
\end{equation}
which is the projection to the normal plane of the curve at $\boldsymbol{x}$; in particular, $\mathsf{f}$ is nonnegative and symmetric.

Applying this force density at $\boldsymbol{x}$ imparts a torque density $\boldsymbol{x} \times (\mathsf{f}\boldsymbol{v})$, and lets us define the associated \emph{torque density matrix}
\begin{equation}
\mathsf{q} = \boldsymbol{x} \times \mathsf{f}.  
\end{equation}
We immediately see that $\mathsf{q}$ is traceless: we view $ \boldsymbol{x} \,\times$ as left multiplication by a skew-symmetric matrix, and note that trace defines an inner product under which the skew-symmetric matrices are perpendicular to the symmetric matrices, like $\mathsf{f}.$

Integrating the force density along the curve with respect to arclength accounts for the total force on the curve 
\begin{equation}
\boldsymbol{F}=\mathsf{F} (\boldsymbol{v}) = \int \mathsf{f}(\boldsymbol{v}) = \int \boldsymbol{v}-\boldsymbol{t}(\boldsymbol{t}^* \boldsymbol{v} )
\end{equation}
which defines the corresponding \emph{total force matrix} 
\begin{equation}
\mathsf{F} = \int \mathsf{f} = \int{\mathsf{I}-\boldsymbol{t}\boldsymbol{t}^*}\hspace{-4pt}.
\end{equation}
Similarly, the \emph{total torque matrix} is given by 
\begin{equation}{\label{totaltorquematrix}}
\mathsf{Q}  = \int \mathsf{q} =  \int{\boldsymbol{x} \times \mathsf{f}} = \int{\boldsymbol{x} \times  \left({\mathsf{I}-\boldsymbol{t}\boldsymbol{t}^*} \right)}.
\end{equation}

\begin{figure}
\centering     
\includegraphics[width=0.45\textwidth]{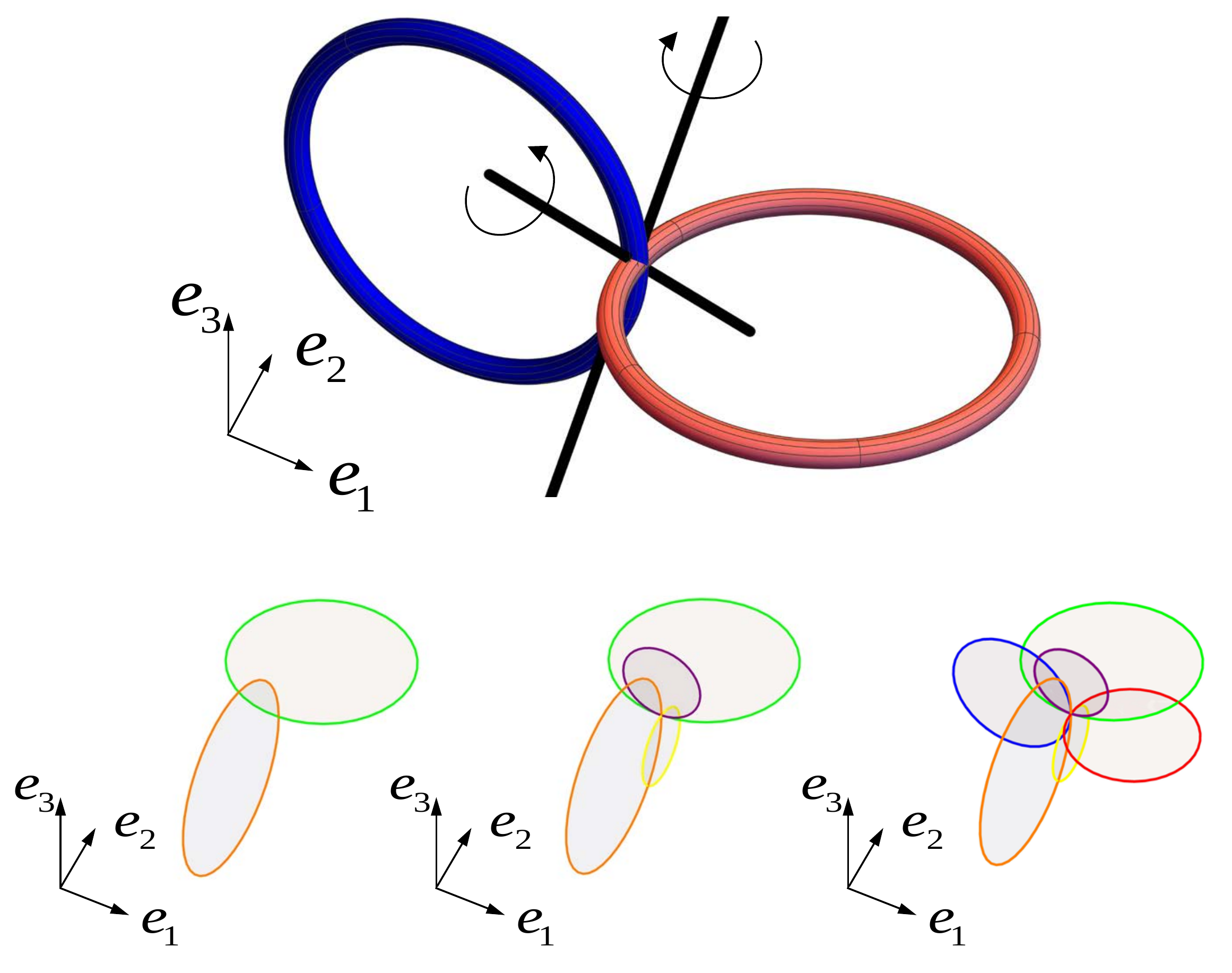}
\caption{A skew propeller composed of two circles in perpendicular planes twists to the right for a wind blowing in the $\boldsymbol{e}_2+\boldsymbol{e}_3$ direction, and to the left for the $\boldsymbol{e}_2-\boldsymbol{e}_3$ direction.  In each case, the axis of the initial rotation is in the direction parallel to the wind. The subfigures illustrate the construction of a set realizing a given torque matrix $\mathsf{Q}$.}
\label{fig4}
\end{figure}

The torque matrix $\mathsf{Q}$ is our proposed tensor measure of chirality for a curve.  
The component ${\boldsymbol{u}}^*\mathsf{Q}{\boldsymbol{v}}$ of the torque $\mathsf{Q}{\boldsymbol{v}}$ induced by a wind with velocity ${\boldsymbol{v}}$ about a line through the origin parallel to $\boldsymbol{u}$ defines a bilinear form. 
The corresponding quadratic form ${\boldsymbol{u}}^*\mathsf{Q}{\boldsymbol{u}}$ measures the {\em twist} of a curve about the unit vector $\boldsymbol{u}$.

The Euclidean group $E(3) = \mathbb{R}^3 \ltimes O(3)$ on $\mathbb{R}^3$ acts by translations and orthogonal transformations
\begin{align}
\boldsymbol{a}_{\star} \boldsymbol{x} &= \boldsymbol{a} + \boldsymbol{x} 
 \hspace{2cm}
 \mathsf{R}_{\star} \boldsymbol{x} = \mathsf{R} \boldsymbol{x},
\end{align}
where $\boldsymbol{a} \in \mathbb{R}^3$ and $\mathsf{R}\in O(3)$. The force and torque densities transform accordingly:
\begin{align}
\label{trans}
\text{(a)} \,\,\boldsymbol{a}_{\star} \mathsf{f} &= \mathsf{f} \hspace{2.27cm}
\text{(b)} \,\,\mathsf{R}_{\star} \mathsf{f} = \mathsf{R} \mathsf{f} \mathsf{R}^*\\
\text{(c)}  \,\,\boldsymbol{a}_{\star} \mathsf{q} &=\mathsf{q} + \boldsymbol{a} \times\mathsf{f} \hspace{1cm}
\text{(d)} \,\,\mathsf{R}_{\star} \mathsf{q} = \mathrm{det} (\mathsf{R}) \mathsf{R} \mathsf{q} \mathsf{R}^*. \nonumber
\end{align}
In particular, $\mathsf{f}$ is translation invariant and $\mathsf{q}$ is transformed to $-\mathsf{q}$ by the action of the antipodal map $\mathsf{R}=-\mathsf{I}$.

The torque matrix depends on a choice of origin for the coordinates: if a point $\boldsymbol{a}$ in $\mathbb{R}^3$  is translated to the origin, by integrating Eq.~\eqref{trans}(c) the torque matrix becomes $\mathsf{Q}_{\boldsymbol{a}}:= \mathsf{Q} - \boldsymbol{a} \times \mathsf{F} $ and remains traceless. 
There is a unique choice of origin ${\boldsymbol{R}}$---analogous to the {\it center of reaction} for viscous flows---for which the resulting torque matrix $\mathsf{Q}_{\boldsymbol{R}}$ is symmetric: its eigenvalues are real and correspond to critical values of ${\boldsymbol{u}}^*\mathsf{Q}_{\boldsymbol{R}}{\boldsymbol{u}}$ viewed as a function on the unit sphere of directions ${\boldsymbol{u}}$ in $\mathbb{R}^3$.  

The eigenvectors of $\mathsf{Q}_{\boldsymbol{R}}$ have a clear physical interpretation (cf.~\cite{HappelBrenner1973}): If a wire object is restrained from either translating or rotating in a wind parallel to a given eigenvector, then the torque on the wire object would be parallel to the wind velocity.  This suggests that each eigenvalue of $\mathsf{Q}_{\boldsymbol{R}}$ can be measured using a torsional pendulum, with a torsion spring parallel to the wind and directed along the corresponding eigenvector.   

Since $\mathsf{Q}_{\boldsymbol{R}}$ is traceless, the sum of its eigenvalues vanishes. 
Thus, in the model adopted here, \emph{there cannot exist an object that twists the same way, independent of the wind direction.} 
Such an object---dubbed {\it isotropic helicoid}---was hypothesized in 1871 by William Thomson, Lord Kelvin \cite{Thomson1871}, who even proposed a possible construction. 
Our model explains why Kelvin's idea has yet to be realized~\cite{Collins2018}. 

%%%%%%%%%%%%%%%%%%%
%
% Subsection B  %\subsection{Examples}
%
%%%%%%%%%%%%%%%%%%%
We now compute the torque matrix for several illustrative examples. 
For a circle of radius $r$ in the $ \boldsymbol{e}_1\boldsymbol{e}_2$-plane centered at the origin, the force matrix is $\mathsf{F} = \pi r (\boldsymbol{e}_1\boldsymbol{e}_1^* + \boldsymbol{e}_2\boldsymbol{e}_2^*+2 \boldsymbol{e}_3 \boldsymbol{e}_3^*)$ and torque matrix vanishes.  
The larger eigenvalue $\mathsf{F}_3$ belonging to $\boldsymbol{e}_3$ reflects the fact that the wind blowing along $\boldsymbol{e}_3$ is always normal to the curve, whereas in the $\boldsymbol{e}_1\boldsymbol{e}_2$-plane the angles of incidence vary between $-\pi/2$ to $\pi/2$, which results in a decrease of the total force.  
Using the transformation laws in Eq.~\eqref{trans} we can derive the force and torque matrix for a skew-propeller made of two such circles that lie in perpendicular planes and touch at a single point (Fig.~\ref{fig4}). 
The force matrix of this curve is
\begin{equation}
\mathsf{F} = \pi r (2\boldsymbol{e}_1\boldsymbol{e}_1^*+3\boldsymbol{e}_2\boldsymbol{e}_2^*+3\boldsymbol{e}_3\boldsymbol{e}_3^*),
\end{equation}
whereas its torque matrix is off-diagonal
\begin{equation}
\mathsf{Q} =  \pi r^2 (\boldsymbol{e}_2 \boldsymbol{e}_3^*+\boldsymbol{e}_3\boldsymbol{e}_2^*)
\end{equation}
meaning the wind directed along $\boldsymbol{e}_2$ triggers an initial rotation about $\boldsymbol{e}_3$ and vice versa.
As a consequence of this computation and the superposition principle, any traceless symmetric $\mathsf{Q}$ can be realized as the torque matrix for the union of three suitably-scaled skew propellers oriented along mutually perpendicular axes. 
In this construction, $\mathsf{Q}$ is represented by a symmetric matrix with diagonal elements all zero.

Next, consider $N$ turns of a helix with radius $r$, pitch $p$ and axis parallel to $\boldsymbol{e}_3$, parametrized by 
\begin{equation}
\boldsymbol{x}(s)=
\left(r \cos u(s), r \sin u(s), p\> u(s)\right)
%\left(r \cos \frac{s}{\sqrt{p^2 + r^2}}, r \sin \frac{s}{\sqrt{p^2 + r^2}}, 
% \frac{sp}{\sqrt{p^2 + r^2}} \right)
\label{hel}
\end{equation}
%with $s \in \left(-N \pi \sqrt{p^2 + r^2},N \pi \sqrt{p^2 + r^2}\right)$.
with $u(s)=\frac{s}{\sqrt{p^2 + r^2} }\in \left(-N \pi ,N \pi \right)$.
Its torque matrix calculated at the center of reaction is given by
\begin{equation}
\mathsf{Q} = \frac{N \pi p r^2}{2 \sqrt{p^2+r^2}} \begin{pmatrix}
    -3       & 0 & 0 \\
    0       & -1 & (-1)^N \frac{2p}{r} \\
     0       & (-1)^N\frac{2p}{r} & 4 
\end{pmatrix}.
\label{pmat}
\end{equation}
Note that $\boldsymbol{e}_1$ is an eigenvector of $\mathsf{Q}$ (with eigenvalue $-3$), which follows from the symmetry of this curve with respect to a rotation by $\pi$ around $\boldsymbol{e}_1$. 
The two other eigenvectors are located in  $\boldsymbol{e}_2\boldsymbol{e}_3$-plane. 
They approach  $\boldsymbol{e}_2$ and $\boldsymbol{e}_3$ as $p/r \rightarrow 0$. 

\begin{figure}
\centering     
\includegraphics[width=0.45\textwidth]{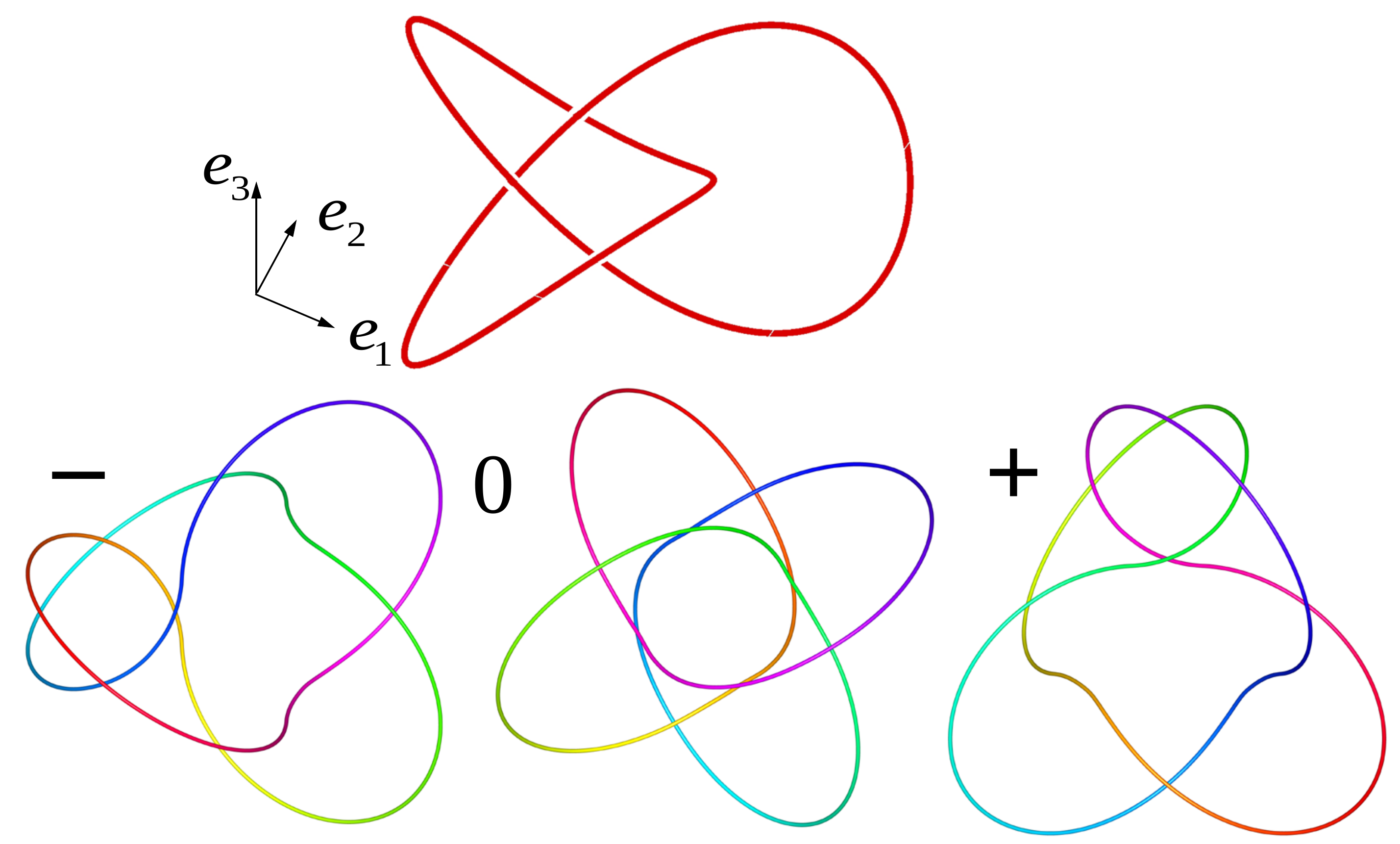}
\caption{Top: A left-handed trefoil is a right-handed propeller. Bottom: Three views along $\mathsf{Q}_{\boldsymbol{R}}$-eigenvectors of a polygonal approximation to the tight figure-eight knot.}
\label{fig5}
\end{figure}

Finally, consider a trefoil (Fig.~\ref{fig5}) parametrized by
\begin{equation}
\boldsymbol{x}(t) = \left(\sin t  + 2 \sin 2t, \cos t  - 2\cos 2t, -\sin 3t\right).
\label{trefoil}
\end{equation}
 
\noindent
Its torque matrix with respect to the origin (the center of reaction in this case) is symmetric and diagonal with two equal eigenvalues, and the third is twice as large with opposite sign:
\begin{equation}
\mathsf{Q} \approx 3.22664 \left(-\boldsymbol{e}_1\boldsymbol{e}_1^*-\boldsymbol{e}_2\boldsymbol{e}_2^*+2\boldsymbol{e}_3\boldsymbol{e}_3^*\right).
\label{trep}
\end{equation}
This is expected from the 3-fold symmetry about $\boldsymbol{e}_3$.

\begin{figure}
\centering     
\includegraphics[width=0.45\textwidth]{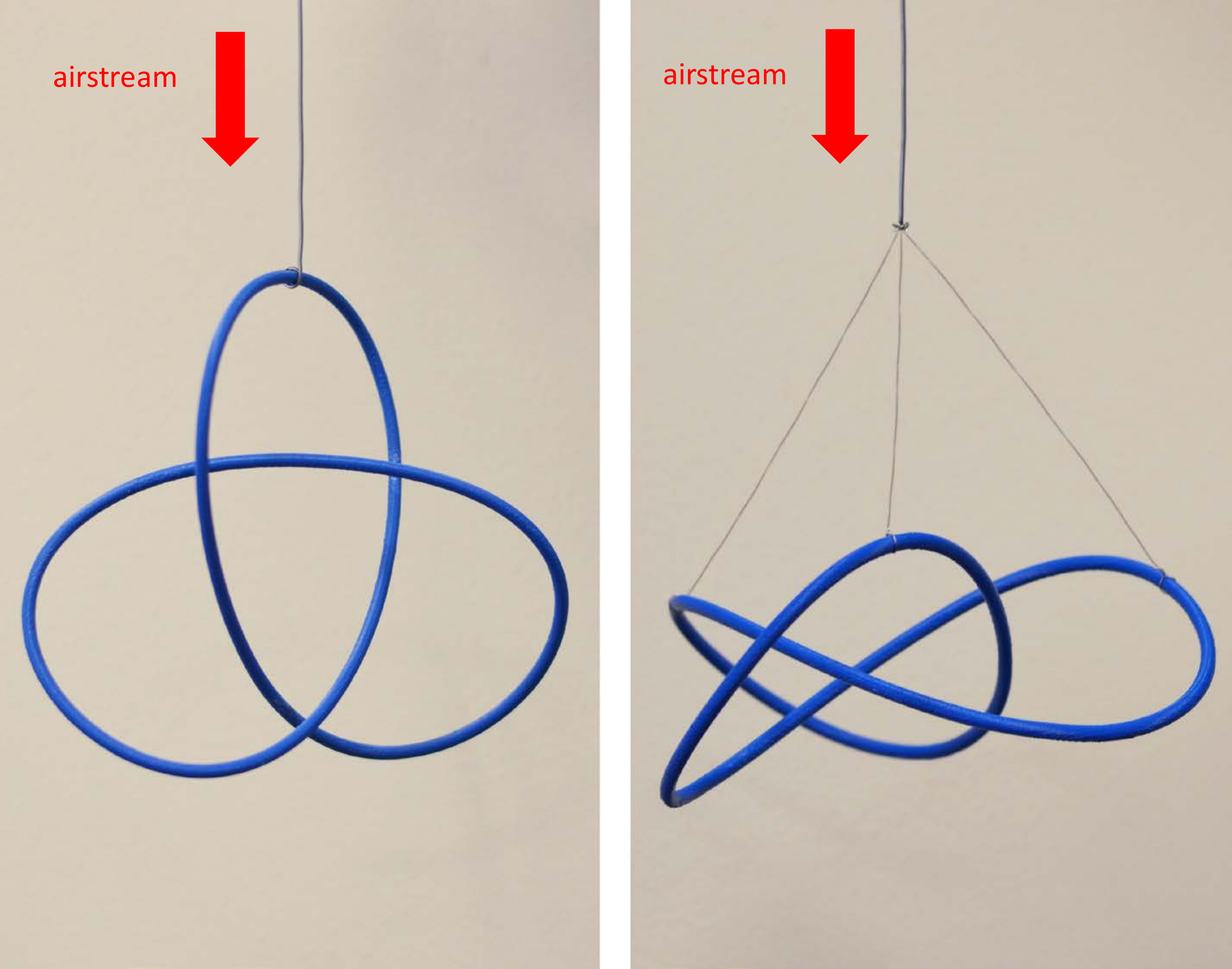}
\caption{The experimental system: trefoil suspended on a torsion wire along $\boldsymbol{e}_1$ (left) and $\boldsymbol{e}_3$ (right). In the first case, the knot rotates in a right-handed manner; in the second case, the rotation is left-handed, with approximately half the magnitude. Image credits: Chris Bartlett and Kristine Henriksen.}
\label{fig6}
\end{figure}

%%%%%%%%%%%%%%%%%%%
%
% Subsection C  %\subsection{Experimental Apparatus}
%
%%%%%%%%%%%%%%%%%%%
To test the predictions of our model, we built a simple experimental system in which a rigid wire object is tethered on a torsion spring and placed in a wind. 
We produced a trefoil by 3D printing a shape given by Eq.~\eqref{trefoil} and tethered along either  $\boldsymbol{e}_1$ or $\boldsymbol{e}_3$ (as illustrated in Fig.~\ref{fig6}). 
The torsion spring has a relatively large stiffness so that the slow rotation of the object can be neglected in the force balance, and yet the total rotation to reach an equilibrium position remains noticeable and can be quantified. 
In the experiment, the trefoil oriented along $\boldsymbol{e}_1$ rotated in a right-handed manner, whereas that oriented along $\boldsymbol{e}_3$ rotated in a left-handed manner, in agreement with the analysis of the eigensystem of the torque matrix in Eq.~\eqref{trep}. 
Moreover, the number of turns required to reach an equilibrium position in the second case was approximately twice as large as in the first case, again consistent with Eq.~\eqref{trep}. 
The agreement of the model with experimental data is reassuring; nevertheless one must remember our model represents a great simplification of the underlying physical process---its motivation was to find a simple way to quantify chirality, rather than a faithful description of the momentum transfer between fluid and object. 

While our experiments are somewhat similar to the experiments of~\cite{Efrati2014}, they differ in conceptual framework. 
In the latter, an object is placed in a wind while rigidly connected to an axis pointing along the wind direction; the wind then induces rotation of the object and the rotational velocity is measured.  
In our case the rotation is constrained by the resistance of a torsion spring. 
In this way we can measure the hydrodynamic torque of the wind acting on the wire object. 
We are essentially solving a friction problem---to compute force and torque on the object for a given velocity---in contrast to the mobility problem considered in~\cite{Efrati2014}. 
Curiously, the measure of chirality proposed in~\cite{Efrati2014} is not directly connected with the translational-rotational coupling in the flow. 
Instead, the entry $Q^{\prime}_{ij}$ of their tensor measure $\mathsf{Q}^{\prime}$ quantifies the rotation of the principal normal about the $\boldsymbol{e}_j$ direction when displaced along the curve in the direction and magnitude projected from $\boldsymbol{e}_i$:
\begin{equation}
Q^{\prime}_{ij}= \int \left( t_i t_j \tau + t_i b_j \kappa \right)
\label{qprime} 
\end{equation}
where $\boldsymbol{b}$ is the binormal vector, $\tau$ is the torsion, and $\kappa$ the curvature. For the trefoil described by Eq.~\eqref{trefoil}, such a chirality matrix is given by
\begin{equation}
\mathsf{Q}^{\prime} \approx 0.480436  \left(\boldsymbol{e}_1\boldsymbol{e}_1^* + \boldsymbol{e}_2\boldsymbol{e}_2^* \right) +1.26417 \boldsymbol{e}_3\boldsymbol{e}_3^*\,.
\label{trep2}
\end{equation}
Note that all the eigenvalues in this case are positive: with respect to their measure in Eq.~\eqref{qprime}, the right-handed trefoil is right-handed when observed from any direction. This is fundamentally different from our torque measure in Eq.~\eqref{trep}, and seems inconsistent with the results of our experiments. 

%%%%%%%%%%%%%%%%%%%
%
% Subsection C  %\subsection{Discussion}
%
%%%%%%%%%%%%%%%%%%%
Perhaps the differing conceptual foundations for $\mathsf{Q}$ and $\mathsf{Q}^{\prime}$ account for this discrepancy.  
Our chirality measure $\mathsf{Q}$ derives from the simple idea of momentum transfer between the wind and the curve, depending only on $C^1$ (tangent) data.  
On the other hand, the measure $\mathsf{Q}^{\prime}$ uses the Frenet frame---not always well-defined---along a curve, and requires more smoothness since its definition in Eq.~\eqref{qprime} depends on $C^2$ (curvature) and $C^3$ (torsion) data; this leads to instability of $\mathsf{Q}^{\prime}$ under perturbations of a curve, and failure to converge under physically natural limits of curves.  
In contrast, our measure $\mathsf{Q}$ has the advantage of being stable under $C^1$ perturbations and limits; furthermore, the smoothness requirements on a curve needed to define our torque matrix in Eq.~\eqref{totaltorquematrix} are relatively mild: \emph{it suffices for the unit tangent vector to exist almost everywhere with respect to the arclength measure}, which holds for rather crooked curves like polygons and more general $1$-dimensional rectifiable sets. 

This feature lets us use polygonal approximations to accurately compute the chirality of curves which do not have a reasonable parametric representation.  For example, the figure-eight ($4_1$) knot is amphichiral---isotopic to its mirror image---and the $\mathsf{Q}_{\boldsymbol{R}}$-eigenvalues $(-1.644854927308887, 5\times10^{-16}, 1.644854927308889)$ for a ropelength-critical configuration (approximated by a polygon with 200 edges and numerically tightened using \texttt{Ridgerunner}\cite{expmath},\cite{cantarella2014symmetric} with four-fold roto-reflectional symmetry enforced: see bottom panel of Fig.~\ref{fig5}) detect the amphichirality of this configuration to high precision. In the above calculation of the eigenvalues of $\mathsf{Q}_{\boldsymbol{R}}$, the radius of gyration of the tight knot was used as a length unit.  Further examples of computations for tight polygonal knots can be found in an online database at \url{george.math.stthomas.edu/~rawdon/Torque/}.

To illustrate why our chirality measure $\mathsf{Q}$ is continuous in the $C^1$-topology on curves, start with the helix defined in Eq.~\eqref{hel} and take a limit $N\to\infty, \, p\to 0, \, r\to 0$ such that $pN=1$ while $ rN \rightarrow 0$. 
This corresponds to a helix winding more and more times about its axis with increasing tightness, whereas the axial extent is fixed.  
The tangent vector 
$$\boldsymbol{t} =\frac{1}{\sqrt{p^2 + r^2}} \left(- r \sin \frac{s}{\sqrt{p^2 + r^2}}, r \cos \frac{s}{\sqrt{p^2 + r^2}}, p\right)$$
 approaches $\boldsymbol{e}_{3}$, since $r/p \to 0$, meaning that the helix $C^1$-converges to an axis segment. An analysis of Eq.~\eqref{pmat} shows that our chirality measure $\mathsf{Q}$ converges to $0$, the chirality of a segment. %..
However, the measure $\mathsf{Q}^{\prime}$ is not $C^1$-continuous: for this helix, the tensor $\mathsf{Q}^{\prime} = \frac{2 \pi N p}{\sqrt{p^2+r^2}} \boldsymbol{e}_3 \boldsymbol{e}_3^*$ diverges in 
the limit where $N\rightarrow\infty$ and $r/p\rightarrow 0$.

The modest smoothness requirements for our simple torque-based chirality measure $\mathsf{Q}$ make it a potentially valuable tool, not only for describing the core curves of classical knots and links in energy-critical configurations, but also for exploring the chiral properties of confined random walks, protein backbones, and polymer chains.  
Furthermore, our approach can be naturally extended to yield a chirality measure for meshes, surfaces, and higher-dimensional objects, and also to other scattering interactions. 

\begin{acknowledgments} 
Special thanks to Keith Moffatt for inspiring the authors to begin thinking together about chirality at the Newton Institute in 2012.  
R.~Kusner was supported in part by a fellowship at the Newton Institute, as well as by NSF grants PHY-1607611 at the Aspen Center for Physics, DMS-1440140 at ICERM, and DMS-1439786 at MSRI.  
W.~Kusner was partially supported by the University of Pittsburgh, Graz University of Technology, Vanderbilt University, Austrian Science Fund (FWF) Project 5503, and NSF grants DMS-1516400 and DMS-1104102.  
E.~Rawdon was partially supported by NSF grants DMS-1115722, DMS-1418869, and DMS-1720342.
P.~Szymczak was supported in
part by a fellowship at the Newton Institute as well as by the National Science Centre (Poland) under research grant 2015/19/D/ST8/03199.  

The authors are listed in alphabetical order.
\end{acknowledgments} 

\bibliographystyle{apsrev4-1}
\bibliography{chirality}

\end{document}